\documentclass[conference]{IEEEtran}

\usepackage{graphicx} 
\usepackage{algorithm}
\usepackage{algorithmicx}
\usepackage{algpseudocode}
\usepackage{amsmath}
\usepackage{array}
\usepackage[caption=false,font=normalsize,labelfont=sf,textfont=sf]{subfig}
\usepackage{soul}
\usepackage{url}
\usepackage{verbatim}
\usepackage{cite}
\usepackage[bookmarks=false]{hyperref}
\usepackage{listings}
\usepackage[table]{xcolor}
\usepackage[nolist]{acronym}
\usepackage{pdfcomment}

\def\BibTeX{{\rm B\kern-.05em{\sc i\kern-.025em b}\kern-.08em
    T\kern-.1667em\lower.7ex\hbox{E}\kern-.125emX}}

\begin{document}

\title{Predictive Intent Maintenance with Intent Drift Detection in Next Generation  Networks\\
}

\author{\IEEEauthorblockN{%
  Chukwuemeka Muonagor, 
  Mounir Bensalem and 
  Admela Jukan} 
\IEEEauthorblockA{%
  Technical University of Braunschweig, Germany \{c.muonagor, mounir.bensalem, a.jukan\}@tu-bs.de}
}

\maketitle

\begin{abstract}

Intent-Based Networking (IBN) is a known concept for enabling the autonomous configuration and self-adaptation of networks. One of the major issues in IBN is maintaining the applied intent due the effects of drifts over time, which is the gradual degradation in the fulfillment of the intents, before they fail. Despite its critical role to intent assurance and maintenance, intent drift detection was largely overlooked in the literature. To fill this gap, we propose an intent drift detection algorithm for predictive maintenance of intents which can use various unsupervised learning techniques (Affinity Propagation, DBSCAN, Gaussian Mixture Models, Hierarchical clustering, K-Means clustering, OPTICS, One-Class SVM), here applied and comparatively analyzed due to their simplicity, yet efficiency in detecting drifts. The results show that DBSCAN is the best model for detecting the intent drifts. The worst performance is exhibited by the Affinity Propagation model, reflected in its poorest accuracy and latency values. 

\end{abstract}

\section{Introduction}
Intent-Based Networking (IBN) has emerged as one of the key technology solutions for autonomous networks \cite{leivadeas2023survey}. IBN is a network management system that makes use of natural language processing to compile humanly understandable commands to the network management system, while using machine learning and artificial intelligence in general, as well as orchestration of network resources to automate the activation of these commands \cite{bensalem2021role}.  An example of intent could be expressed as \emph{permit all traffic from nodes A, F, and K to go through port 80 and not traffic from other nodes in the system}, which expressed in this simple form can assure that the policy in the system is enforced and maintained. According to Cisco \cite{cisco2022intent}, intent assurance is a core part of any IBN system. It is also referred to as intent validation or verification, which is a network management process that continuously monitors the network to verify that the desired intent which has been applied is functioning properly.

\par
\begin{figure}[t]
    \centering
    \includegraphics[width=0.45\textwidth]{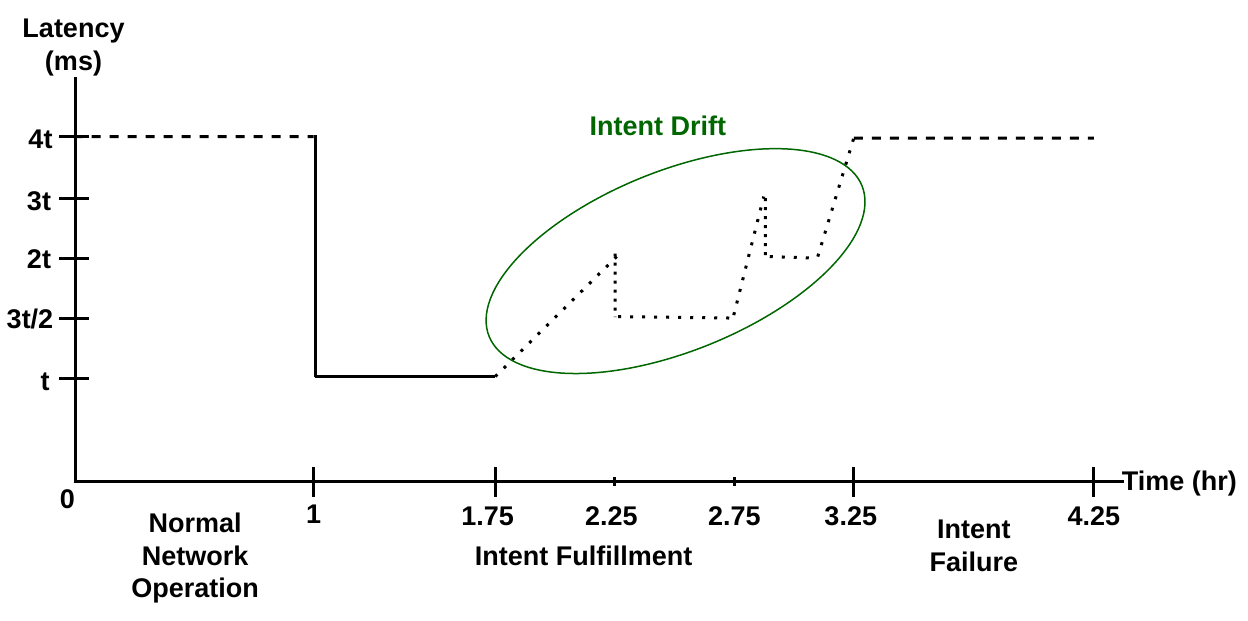}
    \caption{Intent life cycle and the appearance of drifts.}
    \label{fig:lifecycle}
\end{figure}

  A typical intent-based network life cycle considers three phases: Phase 1 -  when the network operates without an active intent; Phase 2 - the period during which the intent is active and the network behaviour complies with the intent; Phase 3 - the intent failure.  Let take an example of average latency as network parameter to be monitored (Figure \ref{fig:lifecycle}).When an intent is entered into the network (Phase 2) after an hour of network operation when the latency was 4t ms (Phase 1)  the latency value is brought down value of 1t ms instead. In Phase 2, which is intent fulfillment period, the latency remains on the average at this value of t ms for about 75 mins before it starts fluctuating gradually in a generally increasing trend. The failure of the intent is confirmed after one and half hours of these fluctuations in Phase 3 - the intent failure period when the network finally goes back to its original state where the latency was 4t ms before the intent was entered. It is precisely in the time period where the latency fluctuating gradually in a generally increasing trend that illustrates the period of the so-called \emph{Intent Drift}. Although a major part of intent maintenance should consider whether intent drift occurs over time \cite{clemm2021draft}, only a few studies addressed intent drift detection. 

This paper proposes a novel function for the predictive maintenance of intents in the IBN architecture, i.e., the intent drift detection. We develop and apply an intent drift detection algorithm that can selectively test seven machine learning (ML) models to detect intent drifts, which we analyze on their ability to predictably  maintain intents based on their resulting accuracy, response time and complexity. Specifically, unsupervised learning techniques (Affinity Propagation, DBSCAN, Gaussian Mixture Models (GMM), Hierarchical clustering, K-Means clustering, OPTICS algorithm, and One-Class SVM) are adapted to this end, while a greedy algorithm is also used for comparison purposes. The results show that DBSCAN has the best accuracy, and the lowest average response time (latency) in detecting intent drifts, though its space complexity is high. The worst detection performance is exhibited by Affinity Propagation algorithm reflected in its poorest accuracy value and maximum latency. To the best of our knowledge, this is the first work to address the problem of intent drift detection, and predictive intent maintenance. 
 
\par The rest of the paper is organized as follows. We present and discuss related work in Section II. Section III proposes the reference IBN architecture with intent drift detection module. Section IV focuses on intent drift detection algorithms. Section V analyzes numerical results. Section VI concludes the paper. 

\section{Related Work}
Previous work majorly focused on intent assurance and measures to overcome failures \cite{leivadeas2023survey, Rothenberg2020intent}. Studies in \cite{khan2020intent,velasco2021end, bensalem2022benchmarking}resorted to ML techniques to build a self-assurance IBN system on top of programmable networks. Alternatively to intent assurance methods, the use of threshold can be adopted, for instance in \cite{nathan2019policy} with rerouting when packet losses exceed the threshold. The problem with setting of thresholds or establishment of specific reference points for the network is, according to \cite{wu2021changes}, that when network behaviour is dynamic, fixed thresholds specifications may not reflect actual network behavior. A drift in the intent may be detected, due to dynamic network state changes, whereby IBN should automatically take actions to revert the network back to compliance state \cite{clemm2020intent}. However, detecting this gradual degradation of intents and at the same time avoiding false positives is a challenge. 

The problem of intent drift can be solved by analyzing the network behaviour during intent fulfillment and normal network operation, which in turn is a problem of statistical or analytical modeling of the network behavior, or based on time series analysis and forecasting, e.g., as in \cite{nguyen2023evaluation}. While statistical modelling may not be best suited to a large amount of data, time-series analysis and rule-based heuristic algorithms also have difficulty scaling \cite{iyer2024predicting}. We opt for unsupervised machine learning here, due to their simplicity, and due to the fact that the network data used in the analyses would come unlabeled. The unsupervised learning models used in our algorithms are \cite{scikit-learn}: Affinity Propagation, DBSCAN, Gaussian Mixture Models (GMM), Hierarchical clustering, K-Means clustering, OPTICS, and One Class SVM. 
\section{Intent Drift Detection}
\subsection{Reference Architecture}

\begin{figure}[t]
    \centering
    \includegraphics[width=0.45\textwidth]{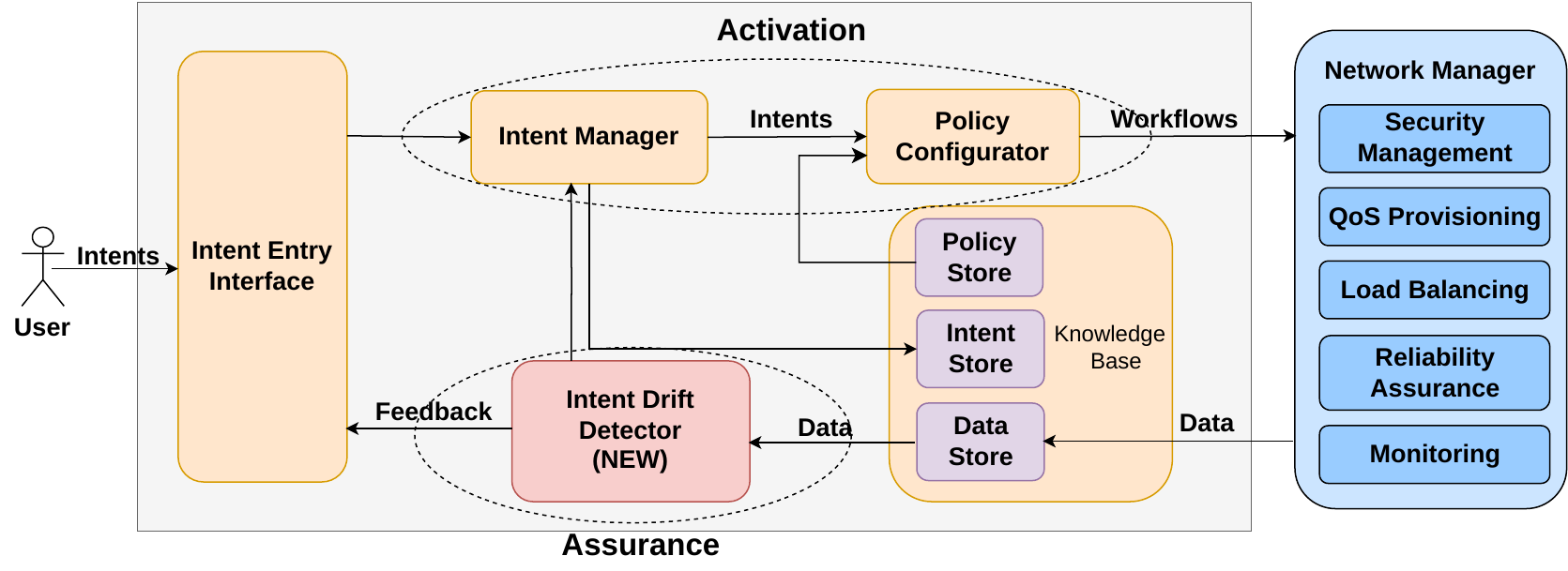}
    \caption{Introducing Intent Drift Detection into the IBN Architecture.}
    \label{fig:XDSM}
\end{figure}

The newly introduced module of Intent Drift Detection into the  typical IBN architecture \cite{clemm2021draft} is illustrated in Figure \ref{fig:XDSM}. Generally, an IBN system fulfills 3 main objectives; \textit{the translation} where the business intents are mapped into policies, \textit{the activation} where the intents and policies are deployed and enforced into the network, and finally \textit{the assurance} where validity of the intent is verified and changes are detected, predicted and prevented, using AI and ML techniques. We also illustrate how IBN interacts with the network management systems, with examples of the selected functions \cite{rafiq2020intent}, including  Security Management, QoS provisioning, etc.

In the IBN architecture, the Intent Entry Interface receives high-level intents from the user and sends same to the Intent Manager, which parses them into low-level intents and forwards them to the Policy Configurator. The Intent Manager also sends the intents to the Intent Store. The Policy Configurator matches the intents received from Intent Manager with the policies in the Policy Store to generate the appropriate workflows to be sent to the Network Manager. The Knowledge Base is the storage component of the IBN system. It contains the Policy, Intent and Data Stores. The Policy Store contains the policies that the Policy Configurator matches the intents with so as to generate the appropriate workflows corresponding to the desired intents. The Intent Store stores the received intents from the Intent Manager. When an intent is deleted from the system, it is also deleted from the Intent Store. The Data Store stores network data received from the Monitoring function block of the Network Manager. 

In this architecture, we propose to add a new component, i.e., the Intent Drift Detector. This  component retrieves data from the Data Store and analyses it to be able to detect any drift in the fulfillment of the intents. It enables predictive maintenance of intents, and prevents its failure.
It should be noted that while the existing IBN architectures consider an intent assurance engine or module, it does not implement any explicit  intent drift detection. This is a novel architectural component introduced in this work which replaces the conventional intent assurance engine. The latter is often hard coded to check periodically the validity of an intent, which is not efficient.

\subsection{Intent Drifts Exemplified}
\begin{figure}[h]
    \centering
    \includegraphics[width=0.45\textwidth]{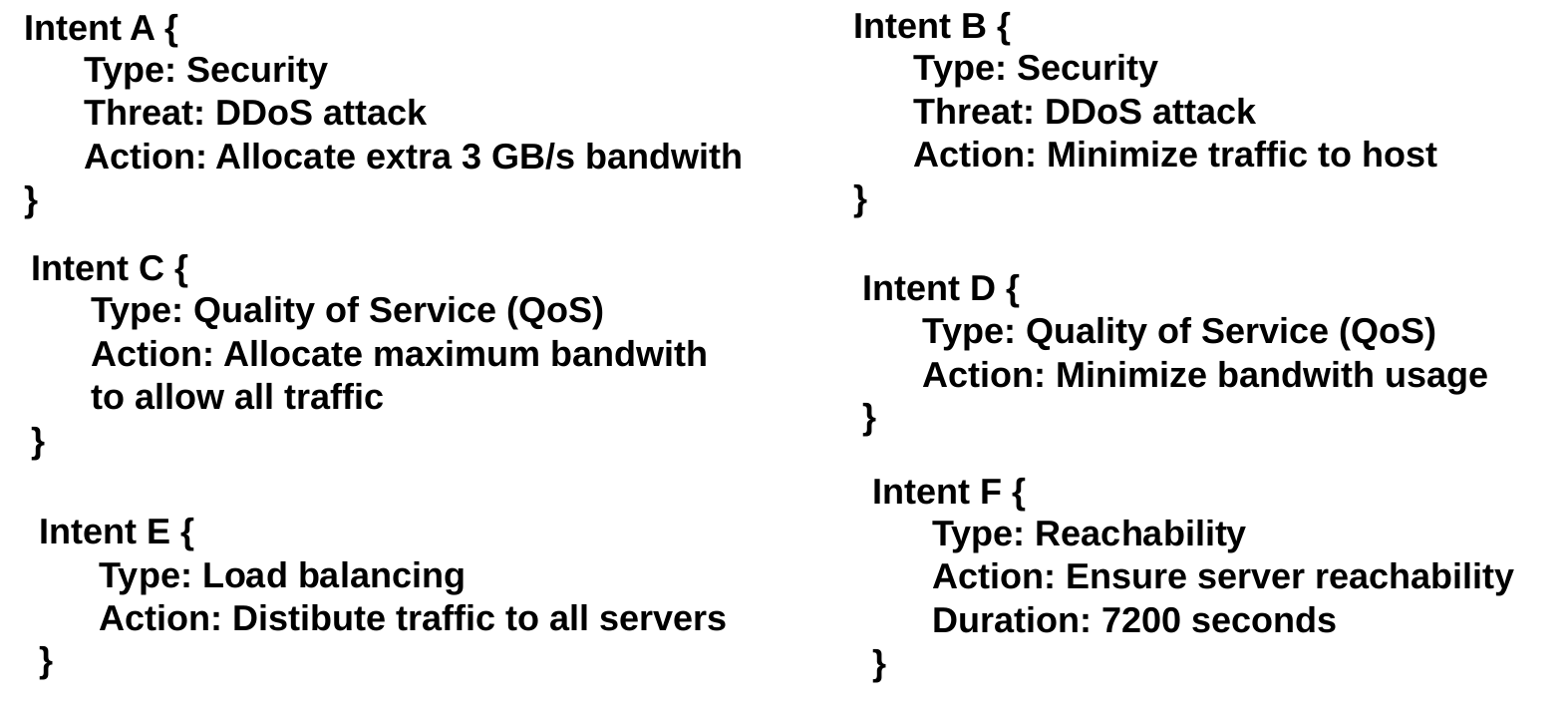}
    \caption{Examples of Intents for Security, QoS provisioning and Reachability}
    \label{fig:intents}
\end{figure}

We now illustrate a few intents and then exemplify their drifts, see Figure \ref{fig:intents}.  Intents A and B are examples of security intents. Intent A entails that the bandwidth allocation to the attack victim is to be increased by an extra 3 Gb/s in the event of the attack. Intent B requires that the entire incoming traffic to the attack victim would be minimised, i.e., many packets would be dropped while very few packets are allowed. Intents C and D are examples of QoS provisioning intents. Intent C implies that there should not be any restriction to the amount of bandwidth allocated to the host. Intent D requires that the bandwidth allocation to all types of traffic to and from each host across the entire network be minimised.  Intent E is a load balancing intent. Finally, Intent F is a reachability assurance intent that ensures that the host or server is reachable for a period of 7200 secs. This implies that all traffic directed to this host must be able to reach it.   

An intent drift can be caused by a  manual intervention of the network administrator: when Intent B is enforced, the Security Management drops several incoming packets to Host  when a DDoS attack is detected. Assuming that an operator manually makes an intervention or changes by allowing some traffic to access the host, this change can cause a drift in the security intent, which needs an additional verification. Another drift example can be caused by an automated reconfiguration: when intent E is enforced, the load balancing distributes the traffic approaching the host. Assuming that the monitoring system can suggest updates to the traffic distribution due to CPU, memory, or bandwidth telemetry, this might lead to unbalanced traffic distribution, which causes intent E to drift. Furthermore, intent drift can be caused by a conflict between intents. For instance, an existing Intent B, and a new intent C can conflict, since due to maximisation of QoS assurance for a host which is what Intent C is, i.e., ensuring the usage of maximum bandwidth by that host, there is conflict with intent B. Similarly, there maybe a conflict between the existing intent A, and new intent D, etc.

\section{Algorithms for Intent Drift Detection}

In this section, we propose an intent drift detection algorithm \textbf{Algorithm \ref{algorithm_1}}, which uses different ML models to detect intent drifts from monitoring time series data. The algorithm receives stored time series data, \textit{ X\_train}, which can be specified depending on the intent, a new incoming data to be evaluated, such as throughput, \textit{ X\_test}, which can be a recently measured telemetry, and finally an algorithm selection, \textit{Model\_Type}, where a list of ML algorithm and heuristics is provided to support drift detection. In \textbf{Step 1}, we check if the chosen model requires a training phase, then run and save the learnt model. In \textbf{Step 2}, we evaluate our new data either using the trained model, or run clustering phase for models like K-means. Using the output of the evaluation phase, the algorithm decides whether there is an intent drift or not depending on how each ML and heuristic works (\textbf{Step 3}), and then send the results to the user (\textbf{Step 4}).

\subsection{Detection algorithms based on various ML models}
\subsubsection{Affinity Propagation}
Affinity propagation creates clusters by sending messages between pairs of samples until convergence. A dataset is then described using a small number of exemplars, which are representative points that have high similarity to other points in the same cluster, and low similarity to points in other clusters. Affinity Propagation has the preference parameter, which is the “availability” a(p, q) sent from point q to point p reflecting how much point q sees itself as an exemplar of point p against other potential exemplars. 
For the detection of intent drift according to Algorithm 1, the multiplier value is initialised at first. Then, the learning process in Step 1 starts with the computation of the suitable preference parameter automatically as the minimum of the pairwise distances between the network throughput values, after which the training is done by generating the number of clusters automatically by assigning samples to exemplars. The higher the variation in the throughput data the higher the number of generated clusters. The same process is repeated for new measurements of throughput values. In Step 3, a drift in the intent is indicated by an increase in the number of generated clusters from the data.
\begin{algorithm}[h]

    \caption{Intent Drift Detection Algorithm}\label{algorithm_1}
    \begin{algorithmic}[1]
        \State{\textbf{Input: }  \textit{X\_train} (time series data, eg. throughput), \textit{X\_test} (new data, e.g,. new throughput values), \textit{Model\_Type} (algorithm type)}
        \State{\textbf{Output: }  boolean (drift detection) }
        \State \textbf{Step 1:}    \If {method requires learning} 
         \State Run learning algorithm of  \textit{Model\_Type} with \textit{X\_train}
         \State Save trained model  
         \EndIf
         
     \State \textbf{Step 2: }  \If {method requires inference} \If {inference uses trained model} 
         \State Evaluate trained model with \textit{X\_test}  
         \EndIf
         \If {inference uses math function} 
         \State Evaluate math function with \textit{X\_test} and \textit{X\_train}
         \EndIf
         \State Save inference results 
        \EndIf
        \State \textbf{Step 3: }  Decide intent drift possibility based on \textit{Model\_Type}   
        \State \textbf{Step 4: }   Send drift detection results to  user.  
%
%
%
    \end{algorithmic}
\end{algorithm}

\subsubsection{DBSCAN}
The intent drift detection process implemented with DBSCAN is based on \cite{scikit-learn} and \cite{putina2021online}. We first initialise MinPts which is the minimum number of samples in a given neighborhood. The learning process starts in Step 1. After the training is done, the neighborhoods are generated automatically based on data variability. The extracted throughput values are then assigned to different neighborhoods or discovered clusters using equation 3. Throughput values that are within the radius of the neighborhood. The more similar the throughput values are, the less the number of generated or discovered clusters or neighborhoods. The same process is repeated for new measurements of throughput values. In Step 3, a drift in the intent is indicated by an increase in the number of generated or discovered clusters from the data. 

\subsubsection{Hierarchical clustering}
Hierarchical clustering builds nested clusters by merging or splitting them successively. The linkage criteria determines the metric used for the merge strategy. 
For the detection of intent drift according to Algorithm 1, the parameters affinity and linkage are initialised at first. The learning process in Step 1 starts with the automatic computation of the ideal distance threshold value which is based on the mean of the extracted throughput data. After which the training is done by generating the number of clusters automatically by assigning samples to clusters. The same process is repeated for new measurements of throughput values. A drift in the intent according to Step 3 can be decided when the number of generated clusters from the data increases. 

\subsubsection{K-Means}
At first, the value of a multiplier, later used in Step 3, is initialised. The input data are the old and new throughput values referred to as X\_train and X\_test, respectively. In Step 1, the learning process is done which involves first determining the optimal number of clusters using the Silhouette analysis since number of clusters determines how accurate the algorithm would be, after which training is done on the extracted throughput data using the determined optimal number of clusters. These throughput values are partitioned into the various clusters and their new centroid values are computed. An array of this new centroid values is created and sorted. The differences between neighbouring centroid values and the maximum of the differences are estimated, the lower this maximum value, the more similar the entire throughput values from different clusters are. The same process is repeated whenever new measurements of the throughput values are received by the model. 
In Step 3, the new maximum is compared with the old maximum. If the new maximum exceeds the product of the old maximum and the multiplier, then there is drift.

\subsubsection{Gaussian Mixture Models (GMM)}
Gaussian mixture model generalises K-Means clustering, also taking note of the input data covariance structure as well as the Gaussian components centroids. One of the important parameters in GMM is the number of desired components. The intent drift detection process using GMM is very similar to that of K-Means. The number of clusters in K-Means is similar to the number of components in GMM and the same process is followed according to the steps in Algorithm 1.

\subsubsection{OPTICS algorithm}
DBSCAN and OPTICS algorithms are similar. For OPTICS algorithm, at first the minimum number of samples, the reachability distance and the minimum cluster size are initialised. Points with similar reachability distances are likely to be in the same cluster. The learning process in Step 1 starts with generating the neighborhoods automatically, assigning the throughput data to different neighborhoods based on their reachability distances. In Step 3, a drift in the intent is indicated by an increase in the number of generated or discovered clusters from the data, after which Step 4 is implemented which reports the intent drift. 

\subsubsection{Support Vector Machine (SVM)}
The SVM algorithm used here is One-class SVM (1-SVM). According to Algorithm 1, the first action is to initialise the kernel function to be used and the nu parameter which is upper bounded by the fraction of outliers considered by the model to be in the data and lower bounded by the fraction of support vectors. This is followed by implementing Step 1. Step 1 involves learning the 1-SVM on the data to group the old extracted throughput values into normal data and outliers. In Step 2, the trained model is used to make inference on the newly extracted data to be able to classify them into either normal data or outliers. The presence of outliers in the newly extracted data indicates possibility of drift, and the drift detection result is sent to the administrator at the controller level.

\subsubsection{Greedy algorithm}
Greedy algorithm is used  for comparison. It is based on the determination of the maximum measured throughput data at any point in time. When the maximum of the incoming traffic over a short period exceeds this network monitoring maximum with a large margin, which is defined at the inception, the intent drift is detected.
 
\subsection{Complexity Analysis}
In terms of data size, aside the GMM and the Affinity Propagation algorithms, the algorithms used here are scalable as they are not affected by the size of data. K-Means, DBSCAN and OPTICS are the most scalable of the models. However, in terms of number of nodes and with a relatively fixed amount of data, the entire algorithms are scalable as they can be deployed on each node and also as many nodes as possible. 

The Affinity Propagation algorithm has a time complexity of the order $O(N\textsuperscript{2}T)$ where $N$ is the number of samples and  $T$ is the number of iterations until convergence. Further, the memory complexity is of maximum $O(N\textsuperscript{2})$. The time complexity of k-means algorithm is $O(N\textsuperscript{2}$). The overall complexity is $O(N\textsuperscript{2})$ for DBSCAN, $O(N\textsuperscript{3})$ for GMM, O(N\textsuperscript) for OPTICS, and $O(N)$ for Greedy algorithm. The complexity of One-Class SVM is at best $O(N\textsuperscript{2})$. The time complexity of the agglomerative hierarchical clustering is $O(N\textsuperscript{3})$ because in each of the $N-1$ iterations the $N \times N$ matrix dist matrix is scanned exhaustively for the lowest distance, and the memory complexity is $O(N\textsuperscript{2})$.

\section{Measurements and Results}
\subsection{Network Emulation Setup and Evaluation Metrics}
We used a network emulation setup which contains three hosts and one virtual switch hosted as virtual machines provisioned with the virtualisation platform, VirtualBox, together with an external host where the VirtualBox platform is hosted. The number of hosts was decided arbitrarily. The external host serves as the controller where the IBN software runs and controls how the rest of the hosts which are the virtual machines operate, and also has the network manager emulator that implements the workflows as commands to the other hosts through \texttt{ssh}. The network data are the ingress throughput at a host. Two iterations each were run on three virtual machines and the performance of the models were measured. The intent drift detection algorithms are implemented on the hosts while they send the results of their analyses to the controller in the external host to be viewed at the intent entry interface.



Since any intent implementation needs to be specific to the intent, we consider two specific intents in the analyses which are applied one after the other. These intents are Intents B and D respectively which have been exemplified earlier. Intent B is an already-existing security and reliability intent which is to deploy an action in the event of an attack so as to mitigate the attack. In the event of the attack the execution of the security intent installed at the controller machine which is to mitigate the attack is triggered. The second intent, Intent D is a QoS intent that entails minimising ingress bandwidth to a host due to load or congestion. This implies that only part of the ingress traffic is selected and routed to the host while the remaining traffic is dropped. For each of the two intents, during the period of the intent fulfillment, the ingress throughput to the host is observed and analysed with the aid of the models to determine when possible drift in the intent starts occurring and/or when the intent fails entirely. The measurement of the ingress throughput is conducted by the host with the aid of the \texttt{ifstat} linux command every second, and the data can be extracted in batches periodically or singularly. Here, an arbitrary choice of every nine seconds is made.

Each measurement iteration contains normal operation, intent fulfillment, intent drift, and intent failure phases. Our focus is on the intent drift phase. The metrics used to judge the performances of the different models are the accuracies of the models observed over the period of the entire iterations, the average latencies of the models while detecting the intent drift periods, and the space complexity of the models.

\begin{figure}[h]
    \centering
    \includegraphics[width=0.45\textwidth]{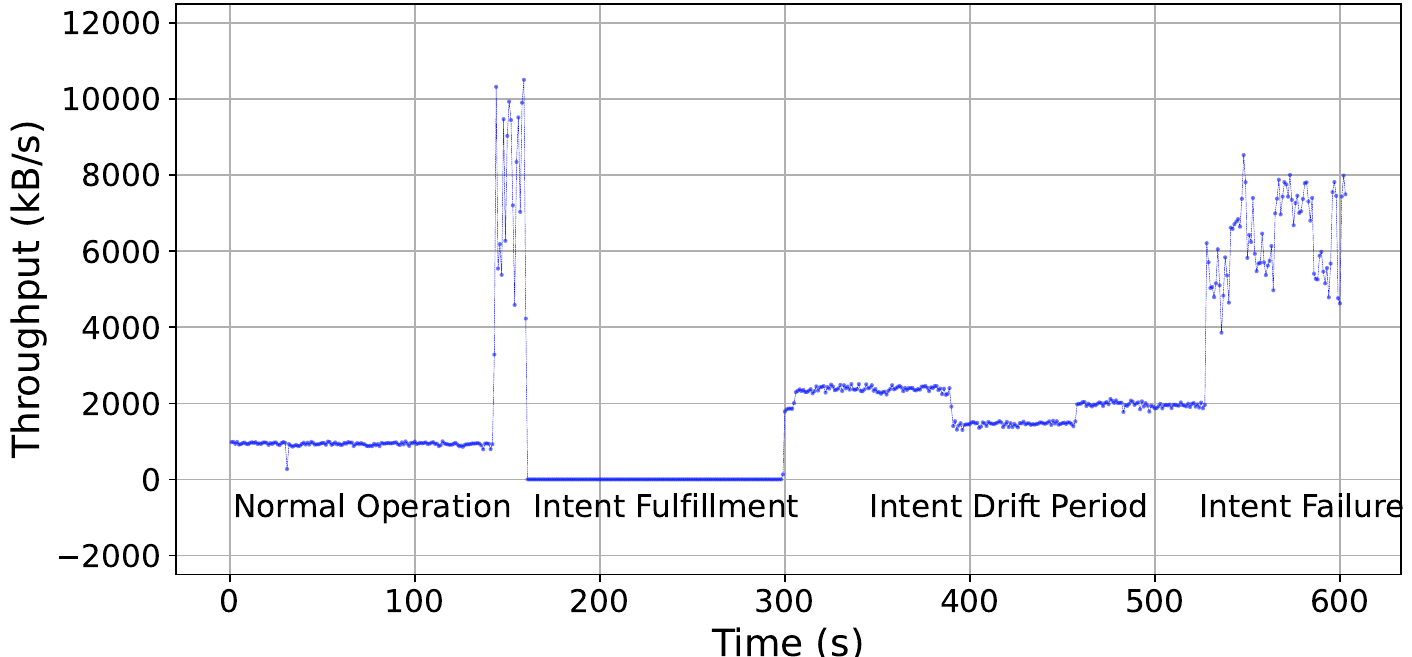}
    \caption{Iteration 1}
    \label{fig:it_1}
\end{figure}

\begin{figure}[h]
    \centering
    \includegraphics[width=0.45\textwidth]{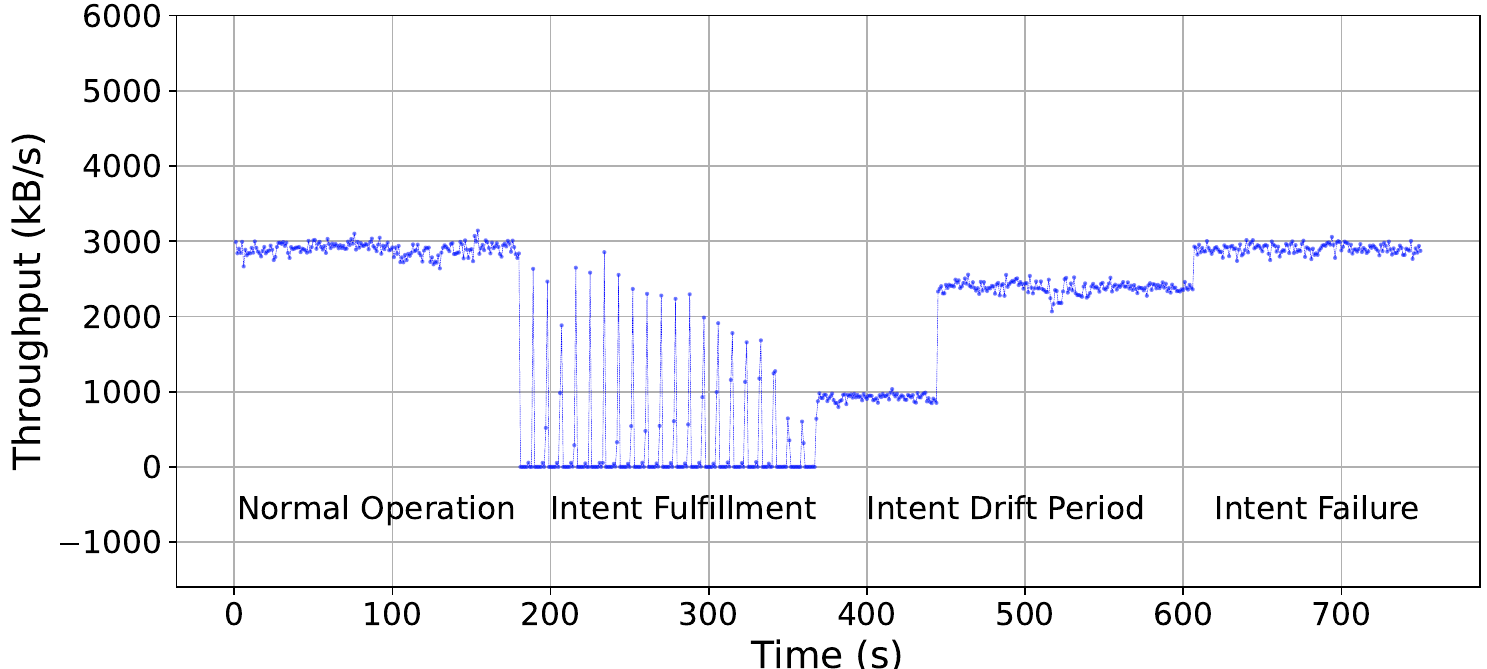}
    \caption{Iteration 2}
    \label{fig:it_2}
\end{figure}

For the two intents, Intents B and D, two of the iterations run on different machines were selected and the plots of the network behaviour capturing their life cycles as illustrated in Figures \ref{fig:it_1} (Intent B - security intent) and \ref{fig:it_2} (Intent D - QoS intent). Figure \ref{fig:it_1} shows that the host experienced an attack at about 140s in time, at which a traffic restriction action according to the existing reliability intent is deployed to mitigate this attack. This precedes the intent fulfillment phase were the attack is mitigated for over 120s due to the action deployed by the intent. This is followed by an intent drift period where the intent fulfillment starts experiencing degradation. This intent drift period lasts for a while before the intent fails eventually when the host starts to experience an attack again. The life cycle for the QoS intent is shown in Figure \ref{fig:it_2}. The intent fulfillment phase starts from almost 180s and lasts for some time before the drift in the intent starts being noticed at almost 400s.

\subsection{Performances of the Models}

\subsubsection{Model Accuracy}
\begin{figure}[h]
    \centering
    \includegraphics[width=0.45\textwidth]{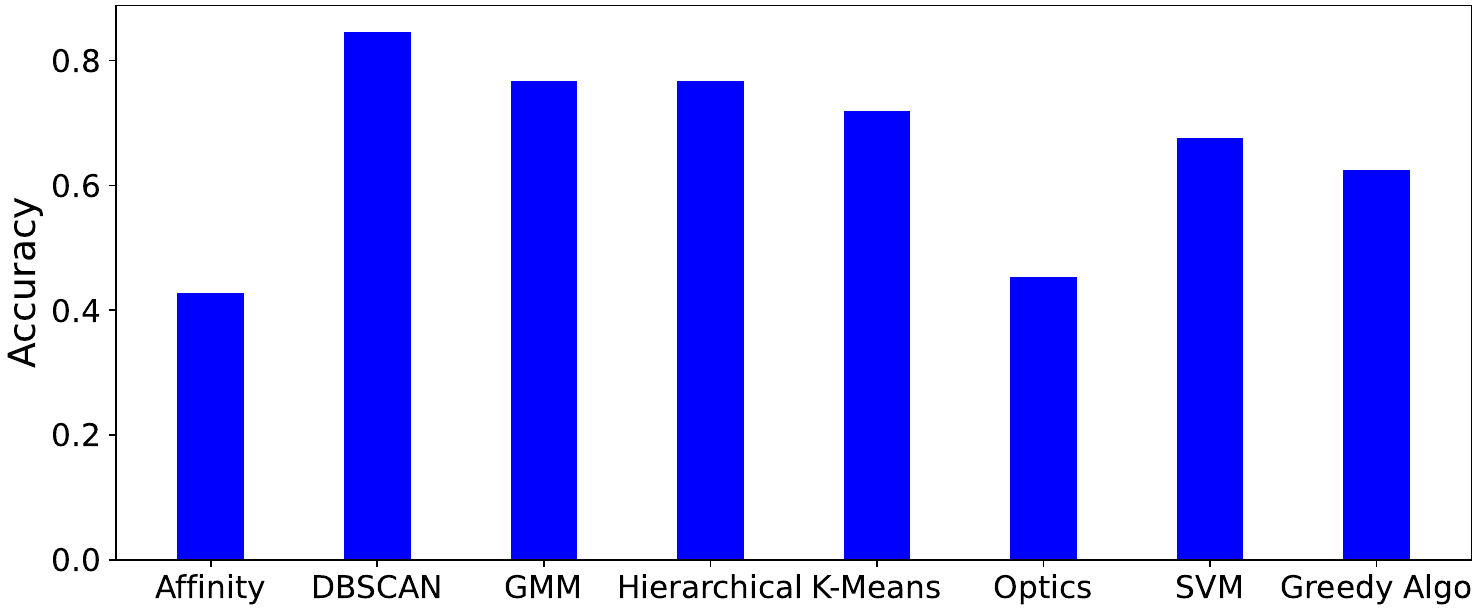}
    \caption{Accuracies of the Models}
    \label{fig:accuracy}
\end{figure}

From Figure \ref{fig:accuracy}, it can be seen that DBSCAN had the highest accuracy (0.846) of all the models in detecting intent drift. This is followed by GMM and Hierarchical clustering methods with a joint accuracy of 0.768, and K-Means clustering, then SVM. The higher accuracy of the DBSCAN model can be attributed to two reasons. One is the fact that in DBSCAN algorithms number of clusters are generated automatically according to statistical dispersion of the data, otherwise known as data variability as opposed to the K-Means, GMM and some other methods that make use of fixed number of clusters which can't always reflect network behaviour or the Greedy algorithm that only computes the maximum values of the throughput data. Another reason is that DBSCAN algorithms are also robust to noise and outliers since they are not assigned to any cluster. The low accuracy of SVM can be attributed to the fact that it tends to suffer when there is no clear separation between the normal and abnormal data, which is the case in this work. 
\subsubsection{Average Latency}
\begin{figure}[h]
    \centering
    \includegraphics[width=0.45\textwidth]{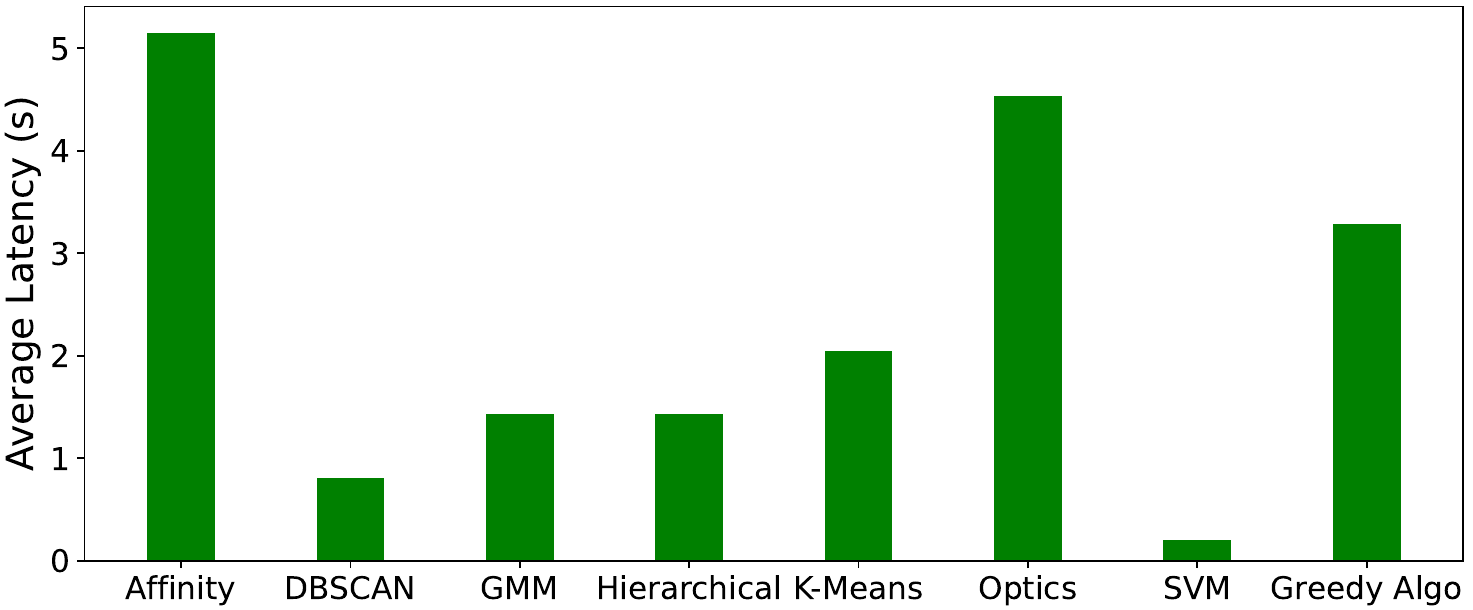}
    \caption{Average Latencies of Models}
    \label{fig:latencies}
\end{figure}

 The resilience and reliability of networks is dependent on the amount of time it takes the network to detect a problem or failure and implement a recovery strategy to ensure resilience. From Figure \ref{fig:latencies}, it is noticed that the SVM model had the lowest average latency in reporting the drifts. In a case like this where the amount of data is not very large, SVM tends to execute fast. Worthy of note is that SVM reported drifts more frequently than any other model even when actual drifts didn't occur, this contributed to its low latency even though it had poor accuracy. DBSCAN algorithm had the second lowest delay while detecting the drift in the network which is followed by GMM. Other algorithms like KMeans, Optics and the Greedy Algorithm showed very high latencies with Affinity Propagation being the poorest due to its general inability to quickly detect the drifts. This can be attributed to the fact that DBSCAN works well with data that are not very well-separated, e.g., network data, and does its clustering as new data are observed unlike K-Means and the others whose strength is with well-separated data, so the former would require less time to draw conclusion that a drift in the intent exists based on the data it observes.

\subsubsection{Space Complexity}
\begin{figure}[h]
    \centering
    \includegraphics[width=0.45\textwidth]{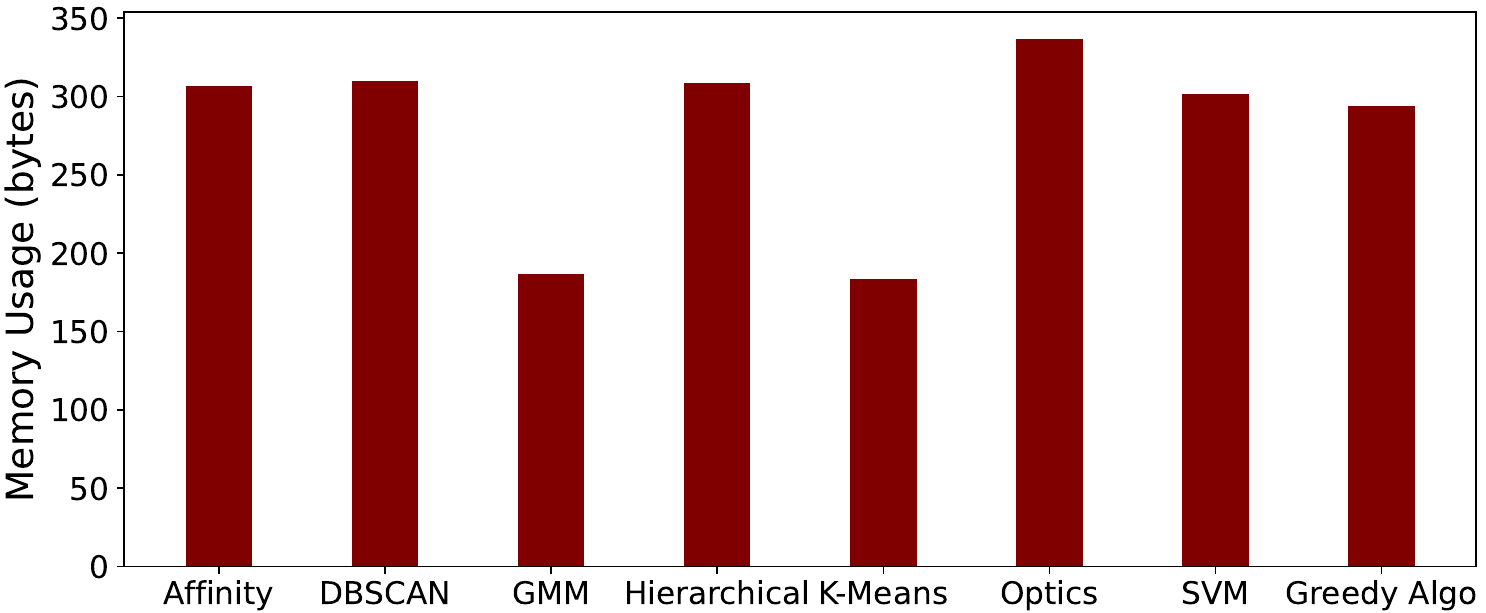}
    \caption{Average Memory Usages of Models}
    \label{fig:memory}
\end{figure}

Space complexity can be estimated as the amount of allocated bytes, whereby algorithms that generally require less spaces run faster. From Figure \ref{fig:memory} which illustrates the average memory usages of the models it can be noticed that OPTICS, DBSCAN, Hierarchical clustering, and Affinity Propagation methods required generally more amount of memory than the others followed by SVM. This high complexity of DBSCAN and the others is due to the fact that they spend time to generate the appropriate clusters themselves automatically before assigning data points to these clusters. GMM and KMeans have the lowest complexity. This is because they view the dataset as well-separated and try to accommodate noise, unlike the previously mentioned algorithms that take strong cognisance of the data variability and do extra work of noticing noise and treating it as outlier.

\section{Conclusions and Outlook}
We conclude that the concept of intent drift detection is very important in predictive intent maintenance.   The DBSCAN model had the best accuracy in detecting the drift due to its automatic cluster generation and tendency to ignore noise and outliers. It also responded second fastest to classifying data traffic as anomaly which is indicated by its lowest average latency of 0.8s, with SVM being the fastest to report a drift. The worst performance was exhibited by the Affinity Propagation model. This is reflected in its poorest accuracy value of 0.428 and maximum latency of 5.152s and also its inability to detect the intent drift phase quick enough. The OPTICS algorithm had the highest complexity while GMM and K-Means had the least complexities.

There are many open areas for further work in this subject. The problem of intent drift is much more complex than illustrated here. Reducing the amount of false positives to the barest minimum is part of the basis for further development in this area of research. Further, the intent drift detection technique is peculiar to the type of intent, so replicating this work for other types of intents in different kinds of network is of very good interest. Also, other methods like neural networks and reinforcement learning can't entirely be ruled out and need to be studied on efficiency. 



\bibliographystyle{IEEEtran}
\bibliography{references_2}

\end{document}